\documentclass{article}
\usepackage{emulateapj,psfig}

\def\cf{cf.}
\def\eg{{\it e.g.}}

\def\etal{{\it et al.}}
\def\kms{\ensuremath{\mbox{km}\,\mbox{s}^{-1}}}
\def\kmsmpc{\ensuremath{\mbox{km}\,\mbox{s}^{-1}\,\mbox{Mpc}^{-1}}}

\def\simlt{\mathrel{\spose{\lower 3pt\hbox{$\mathchar"218$}}
     \raise 2.0pt\hbox{$\mathchar"13C$}}}
\def\simgt{\mathrel{\spose{\lower 3pt\hbox{$\mathchar"218$}}
'     \raise 2.0pt\hbox{$\mathchar"13E$}}}
\def\gsim{ \lower .75ex \hbox{$\sim$} \llap{\raise .27ex \hbox{$>$}} }
\def\lsim{ \lower .75ex \hbox{$\sim$} \llap{\raise .27ex \hbox{$<$}} }

\accepted{}

\long\def\***#1{{\scshape ***#1***}}
\def\arraystretch{1.25}

\makeatletter

\newenvironment{inlinefigure}{%
\def\@captype{figure}%
\noindent\begin{minipage}{0.999\linewidth}\begin{center}}
{\end{center}\end{minipage}\smallskip}
\makeatother

\begin{document}
\lefthead{Moore et al.}
\righthead{Collisional dark matter}
\submitted{Submitted to ApJ Letters, February 16th, 2000}

\title{Collisional versus collisionless dark matter} 
  \author{Ben Moore\altaffilmark{1}, Sergio Gelato\altaffilmark{1}, 
Adrian Jenkins\altaffilmark{1}, 
F. R. Pearce\altaffilmark{1} 
\& Vicent Quilis\altaffilmark{1}}

\altaffiltext{1}{Physics Department, University of Durham, Durham City, UK}

\begin{abstract}

We compare the structure and substructure of dark matter halos in
model universes dominated by collisional, strongly self interacting
dark matter (SIDM) and collisionless, weakly interacting dark matter
(CDM). While SIDM virialised halos are more nearly spherical than CDM
halos, they can be rotationally flattened by as much as 20\% in their
inner regions.  Substructure halos suffer ram-pressure truncation and
drag which are more rapid and severe than their gravitational
counterparts tidal stripping and dynamical friction. Lensing
constraints on the size of galactic halos in clusters are a factor of
two smaller than predicted by gravitational stripping, and the recent
detection of tidal streams of stars escaping from the satellite galaxy
Carina suggests that its tidal radius is close to its optical radius
of a few hundred parsecs --- an order of magnitude smaller than
predicted by CDM models but consistent with SIDM.  The orbits of SIDM
satellites suffer significant velocity bias $\sigma_{_{\rm
SIDM}}/\sigma_{_{\rm CDM}}=0.85$ and are more circular than CDM,
$\beta_{_{\rm SIDM}} \approx 0.5$, in agreement with the inferred
orbits of the Galaxy's satellites. In the limit of a short mean free
path, SIDM halos have singular isothermal density profiles, thus in
its simplest incarnation SIDM is inconsistent with galactic rotation
curves.

\end{abstract}


\section{Introduction}

The nature of dark matter is still far from being resolved. Primordial
nucleosynthesis and observational data suggest that the baryonic
material accounts for just a fraction of the matter density in the
universe.  Fundamental particles remain the most likely candidate for
the dark matter and much effort has been devoted to
researching a class of weakly interacting, collisionless dark matter
(CDM) (\eg, Davis \etal{} 1985). However, the hierarchical gravitational
collapse of cold collisionless particles leads to dense, 
singular dark matter halos -- a result that is central to several 
fundamental problems with this model on small scales 
(\eg~Hogan \& Dalcanton 2000 and references within).

It may be possible to solve the current problems with CDM by appealing
to extreme astrophysical processes. Alternatively, we can explore
other dark matter candidates that behave differently on non-linear
scales. One possibility is strongly self interacting dark matter
(hereafter SIDM). Originally proposed to suppress small scale power in
the standard CDM model (Carlson \etal{} 1992, Machaceck \etal{} 1994, de
Laix \etal{} 1995), SIDM was recently revived by Spergel \& Steinhardt
(1999) to solve some of the outstanding problems with CDM.  The
behaviour of this component depends on the particles' collisional
cross-section. Large cross-sections imply short mean free paths, so
that the dark matter can be described as a fluid that does not cool
but can shock heat. Particles with a mean free path of order the scale
length of a dark matter halo offers the possibility of conductive heat
transfer to the halo cores (Spergel \& Steinhardt 1999).  In this
Letter we contrast the dynamics and structure of ``halos within
halos'' between collisional and collisionless dark matter and compare
predictions with current observational constraints.

\

\section{Simulating the structure of SIDM halos}

In this section we present the first numerical calculations of the
structure of dark matter halos in which the particles have a large
interaction cross-section.  Self interacting dark matter behaves like
a collisional gas and its evolution can be simulated using standard
computational fluid dynamics techniques. We model the collisional dark
matter fluid by approximating its behaviour as an ideal gas where the
ratio of specific heats is 5/3. We use the smoothed-particle
hydro-dynamics (SPH) code Hydra (Couchman \etal{} 1995) to follow the
hierarchical growth of a massive dark matter halo.  For added
confidence in the robustness of key results, we perform independent
collapse tests using an evolution of the Benz-Navarro SPH code
(\cf~Gelato \& Sommer-Larsen 1999). 

\begin{figure*}
\centerline{\psfig{figure=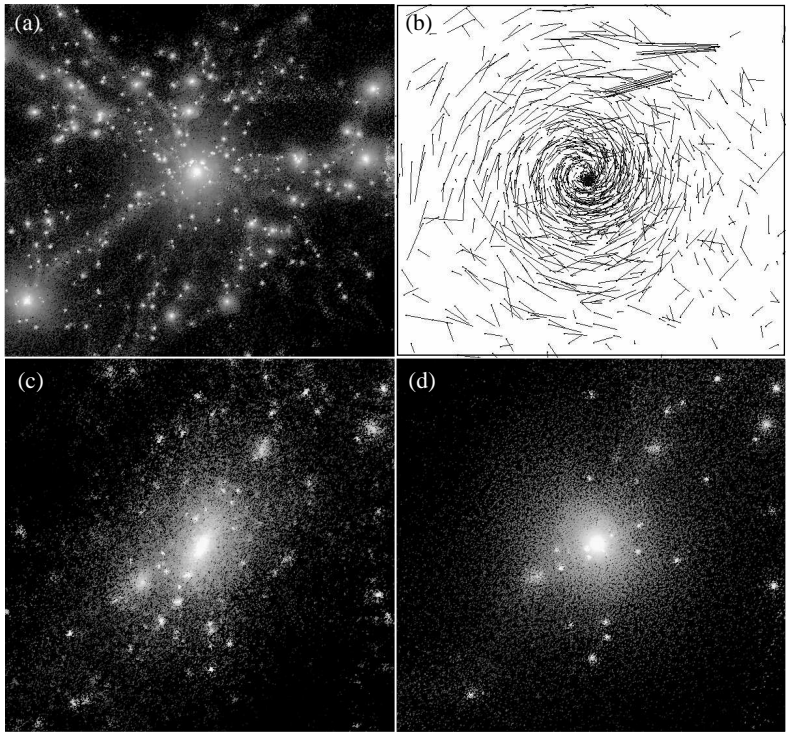,height=0.4\textheight}}
\caption{
Panel (a) is a view of the central 32 Mpc region of our SIDM
simulation at a redshift z=0.  
Panel (b) shows the velocity vectors of particles within the central 500
kpc of one of the SIDM halos. The system is viewed face on to the most
flattened plane demonstrating the coherent rotation.
Dense, but heavily stripped substructure halos can be found sinking deep
into the central regions - frequently prograde with the halo angular momentum. 
Panel (c) shows the high resolution halo in the CDM simulation to be compared
with panel (d) that shows the corresponding halo from the SIDM simulation.
Both of these halos are plotted to their virial radius.
}
\label{fig:a:opt}
\end{figure*}

Our cosmological initial conditions were adapted from the ``cluster
comparison'' simulation (Frenk \etal{} 1999), in which a massive dark
matter halo forms within a 64 Mpc box of a critical density universe.
(We adopt $H_0=50\,\kmsmpc$ throughout.) We carry out
two simulations; the first is a CDM plus 10\% non-radiative gas and
the second run is 100\% non-radiative gas ($\equiv$ SIDM).  The
particle mass is approximately $8.6\times 10^9M_\odot$ and the
effective force resolution is 0.3\% of the virial radius of the final
cluster $r_{\rm vir}=2.7$ Mpc (see Figure~\ref{fig:a:opt}).

The CDM run behaves as expected and as characterised
by many previous authors (\eg, Barnes \& Efstathiou 1987, Frenk \etal{}
1999). One interesting point to highlight from this and similar
simulations is that the gas ends up with a shallower density profile
than the dark matter (\cf~Figure~\ref{fig:b:opt}). 
This is due to energy transfer
between the two components and the fact that the entropy of the gas
can increase through shocks that occur during the gravitational
collapse.  On large scales the SIDM run is similar to the CDM run
although we note that the filaments appear narrower. On non-linear
scales the two models behave very differently and we now discuss the
salient features in more detail.

\subsection{Density profiles}

The final density profile of the most massive SIDM halo is shown next 
to its collisionless counterpart in Figure~\ref{fig:b:opt}. This halo has
more than $10^5$ particles within its virial radius.
The profile is close to a
singular isothermal sphere with slope $\rho(r) \propto r^{-2}$, even
in the very central region.  The hierarchical collapse imparts
thermal energy into the particles which leads to a small amount of
pressure support, however this is not sufficient
to flatten their inner profiles.

To check these results we performed 3D spherical collapses of
power-law spheres with zero initial kinetic energy and density
profiles $\rho(r)\propto r^n$ with $n=-1, 0, +1$. 
We found consistent results with 100 and 5000 particles,
indicating that the singular profile in the cosmological SIDM
simulation is \emph{not} purely an artifact of the high-redshift
progenitor collapses being inadequately resolved.
The collapse with $n=-1$ leads to a singular
spherical isothermal structure.  In this case the central particles
are not strongly shocked and stay at a low entropy.  The $n=0$ and
$n=+1$ collapes generate much higher entropies throughout the
system.  SIDM particles fall in from larger radii achieving higher velocities
and significant thermal energy is generated during the collapse
resulting in a pressure supported constant density core. 
A similar point has been made by Bertschinger (1985). Our
cosmological Gaussian flucuations resemble the former collapse
which results in singular isothermal structures --- what is needed is
a mechanism that prevents low entropy material surviving, such as we
find in more violent collapses.

\subsection{Ram pressure truncation and viscous drag}

Halos of SIDM suffer ram-pressure truncation and ram-pressure/viscous
drag, however dynamical friction is largely suppressed in SIDM models
since the bow shocks and the collisional nature of the fluid 
inhibit the formation of trailing density wakes.
A good approximation is to adopt isothermal profiles for the
substructure halo (subscript $s$) and parent halo (subscript $p$) such
that $\rho(r)=v^2/(4\pi Gr^2)$.  The ram pressure, $\rho(r_p) v_p^2$,
can be equated to the force required to retain a shell of material at
radius $r_s$ from the centre of the substructure halo $F \approx
m_sv_s^2/r_s$. Thus the stripping radius at position $r_p$ in the parent
halo is $r_{\rm strip}=k r_p(v_s/v_p)^2$ where $k$ is a constant of order
$\pi$. This can be contrasted with the tidal radius of embedded
isothermal halos, $r_{\rm tidal}=r_p (v_s/v_p)$.  Therefore, substructure
halos of SIDM will be stripped to substantially smaller sizes than
their CDM counterparts.

It is also interesting to compare the timescale for a substructure
halo to sink to the centre of a larger system due to hydro-dynamical drag,
$F_{\rm drag}=\rho(r_p) v_p^2 4\pi r_s^2$. For a circular orbit, $L=r_p v_p$,
and the rate of specific angular momentum loss, $dL/dt = r_p F/m_s$, therefore
$r_p^{-1} dr_p/dt = F/(m_s v_p)$.  As the substructure is dragged
deeper into the central potential, its radius decreases as calculated
above and we can substitute for $v_s$. Thus we find $dr_p/dt=k v_p$ such
that the drag timescale is simply of order of the crossing time
$t_{\rm drag}=k
r_p/v_p$. All SIDM substructure halos sink at a similar rate
independent of their mass and on a timescale that is typically faster
than that due to dynamical friction.

\begin{inlinefigure}
\centerline{\includegraphics[width=1.0\linewidth]{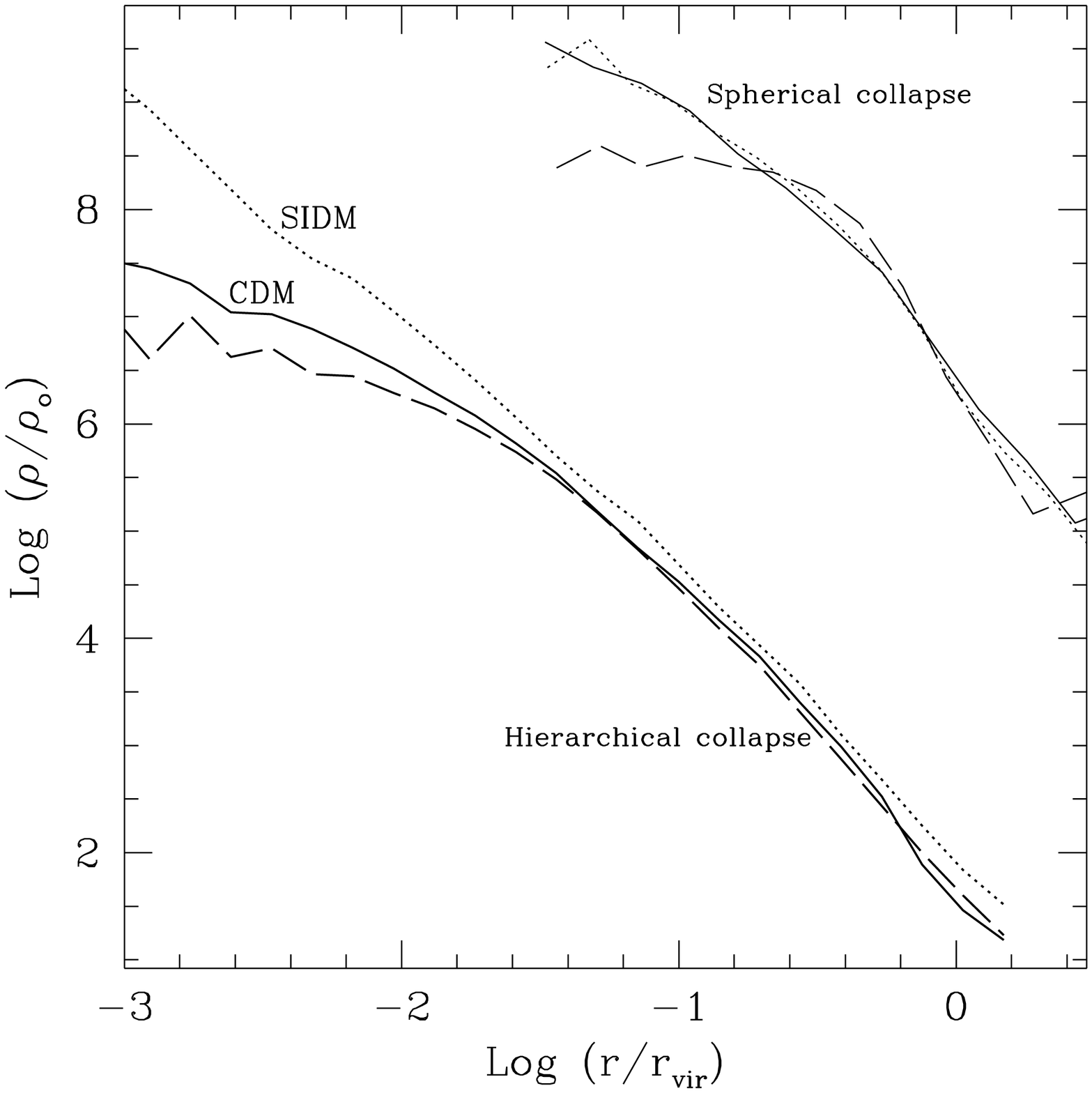}}
\caption{
The radial density profiles of CDM versus SIDM halos. The spherical
collapse halos are plotted using an arbitrary scale at the upper
right. In this case, the solid curve is the CDM density
profile, the dotted and dashed curves are the the spherical SIDM collapses
with $n=0$ and $n=-1$ respectively.  The density profiles of the
hierarchical collapses are shown for SIDM (dotted curve) and for CDM
(solid curve) with 10\% non-radiative gas (dashed curve).
}
\label{fig:b:opt}
\end{inlinefigure}

\subsection{Orbital and velocity bias}

These results have fascinating implications for biasing and the
survival of substructure within dense environments.  In dynamically
old objects, such as galaxy halos, there may have been time for most
of their substructure to sink to the centre. Any surviving
substructure that passes close to the Galactic disk will be stripped
to a negligible mass, therefore disk heating is not a problem in SIDM
models.  Hydro-dynamical destruction may be happening to the
Sagittarius dwarf right now: its current SIDM halo radius would be
approximately 100~pc. We may also expect that galaxies orbiting
through the central regions of rich clusters will have lost most of
their dark matter halos.  Younger systems, such as galaxy clusters
have only had sufficient time to concentrate and bias their
``satellites'' towards the central regions. SIDM satellites suffer
significant velocity bias due to drag: an analysis of the 20 most massive
satellites within the largest dark matter halo yields
$\sigma_{_{\rm SIDM}}/\sigma_{_{\rm CDM}}=0.85$.

The orbits of the Milky Way's satellites with known
proper motions are surprisingly circular (\eg,
Grebel \etal{} 1998,  van den Bosch \etal{} 1999), 
whereas circular orbits are rare in CDM models (Ghigna \etal{}
1998). We find that the anisotropy parameter for SIDM satellites,
$\beta_{_{\rm SIDM}}=0.5$, compared with $\beta_{_{\rm CDM}}=0.32$ (where 
$\beta=v_t^2/(v_t^2+v_r^2)$), which
results from the efficient angular momentum loss of satellites at
pericentre.  SIDM may also account for the ``Holmberg--Zaritsky''
effect (Holmberg 1969, Zaritsky \etal{} 1997). The angular momentum of
SIDM halos is re-distributed differently than in the CDM halos
leading to a rotationally flattened central core.
The baryons are most likely to dissipate into this plane
that aligns with the large scale filamentary structure. It is
material that flows from these cold filaments into the larger halos
that spins up the dark matter: satellites infalling along this
``special'' plane will rapidly sink once they make contact with the
SIDM galaxy halos.  Furthermore, those satellites sinking in the
retrograde direction to the parent halo's angular momentum will be
preferentially destroyed due to the enhanced drag which is $\propto v^2$
(\cf~Figure~\ref{fig:a:opt}b).

\subsection{Halo shapes}

The shapes of dark matter halos provide another clear discriminant
between SIDM and CDM.  The typical ratio of short to long axis for 
CDM halos is
0.5 with a log-normal distribution (Barnes \& Efstathiou
1987). Figure~\ref{fig:c:opt} 
shows the ratio of short to long axis, $c/a$, and
intermediate to long axis, $b/a$, as a function of radius for a well
resolved halo in the simulation.  The virialised part of the halo is
rotationally flattened into an oblate shape such that $\epsilon_{\rm max}
\approx 0.2$.  This is typical of the other SIDM halos which are generally
flattened in the range $0.0\lsim \epsilon \lsim 0.2$.  For comparison
we also show the shape of the same halo in the collisionless CDM
simulation which has a prolate configuration with $c/a \approx
b/a=0.6$ within $r_{\rm vir}$.

\begin{inlinefigure}
\centerline{\includegraphics[width=1.0\linewidth]{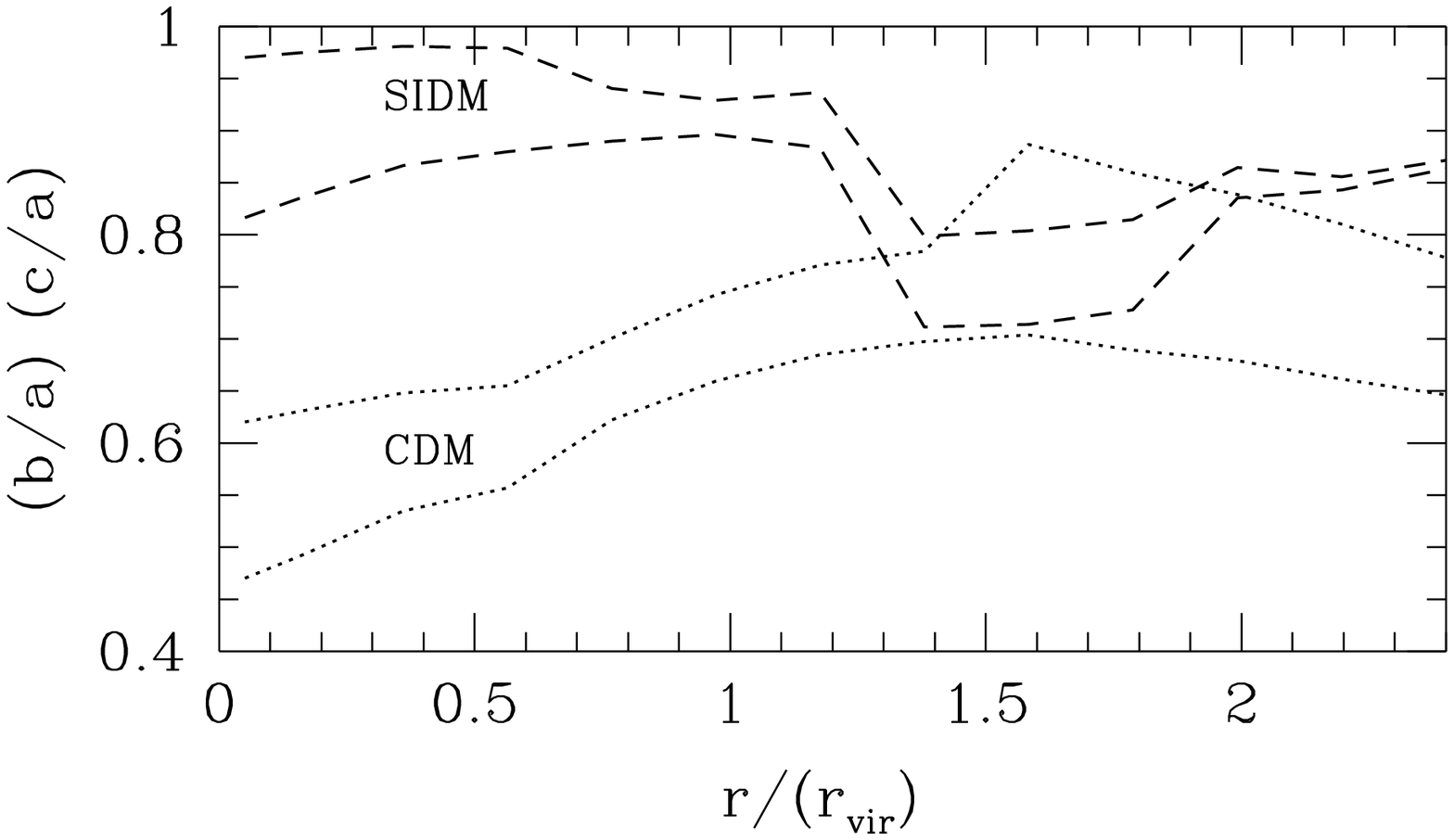}}
\caption{
The axial ratios of a CDM and an SIDM halo are plotted as a function of
radius from the centre. Within the virial radius, this CDM halo is
prolate, whereas the SIDM halo is slightly flattened by
rotation into an oblate configuration.
}
\label{fig:c:opt}
\end{inlinefigure}

Analyses of polar ring galaxies and X-ray isophotes tend to give
flattened dark matter potentials, whereas techniques that use disk
flaring and the precession of warps yield spherical mass distributions
(Olling \& Merrifield 1998).  Ultimately, gravitational lensing
will resolve this issue, but for now we note that a lensing study of
CL0024+1645 constrains the assymetry of the projected mass
distribution to be less than 3\% (Tyson \etal{} 1998). With the notion
that collisional halos should be spherical, Miralda-Escude (2000)
argued that the cluster MS2137-23 rules out SIDM since analysis of its
gravitational arcs demonstrates that its mass distribution must be
flattened such that $\epsilon \gsim 0.1$ in the central region. 
At the moment, SIDM and CDM are both consistent with these data.

\subsection{The extent of halos within halos}

The dwarf satellites of the Milky Way have internal velocities of
order 10--30~\kms, that in isolation would extend to 10--30~kpc
but are tidally limited according to their orbits within the
Milky Way's potential.  Numerical simulations confirm this simple
expectation (Ghigna \etal{} 1998).  For example, the dark matter halo
surrounding the Carina satellite would be truncated to $r_{\rm tidal}
\approx (r_{\rm peri}/50\,{\rm kpc}) (v_{_{\rm Carina}}/v_{_{\rm MW}}) = 2.7\,{\rm kpc}$
at its current position.  In an SIDM universe, the halo of Carina would
be reduced to a size $r_{\rm strip} \approx 400\,{\rm pc}$.

Observations of stars escaping from satellites constrain the extent
of their dark matter halos (Moore 1996, Burkert 1997). Tidal
streams have recently been spectroscopically confirmed for Carina
(Majewski \etal{} 1999) and are also claimed for Draco and Ursa~Minor
(Irwin \& Hatzidimitriou 1993).  These observations imply that the
dark matter extends only as far as the optical radii, about 300
parsecs for all of these satellites and much smaller than their
expected sizes if they had halos of CDM. 

Similarly, the dark matter halos of cluster galaxies are truncated by
the global cluster potential and their sizes can be constrained by
quantifying their effects on strongly and weakly lensed images of
background galaxies.  Natarajan \etal{} (1999) have analysed several of
the clusters imaged by the Hubble Space Telescope and claim that the
dark matter halos of bright cluster galaxies are severely truncated to
between 15--30 kpc. These galaxies have typical internal velocity
dispersions of $150\,\kms$ and sample the projected central 500~kpc
region of the clusters (else they wouldn't lie in the HST frames). 
Thus we expect $r_{\rm tidal}\approx$~30--60~kpc from gravitational stripping, but
$r_{\rm strip}\approx$~10--30~kpc from maximal collisional stripping.

\section{Discussion}

The properties of dark matter halos of strongly interacting particles
are markedly different from their collisionless counterparts. SIDM
halos are close to spherical with a modest degree of rotational
flattening.  Observations of halo shapes cannot currently distinguish
between the models examined here; however, future lensing observations
will determine if SIDM is a viable dark matter candidate.  Halos
within halos suffer ram-pressure truncation that decreases their sizes
to less than the tidal radius. Current observational data on
galactic halos in clusters and satellite galaxies in the Galactic halo
are naturally reproduced in SIDM models: the extent of Carina's halo is
an order of magnitude smaller than predicted by CDM.  Ram-pressure
drag creates significant velocity and orbital bias in the substructure
halos which sink on a short timescale---of order the crossing
time---independent of their mass. Another positive feature of SIDM is
the ability to
produce satellite systems on near circular orbits which are very rare
in CDM models.

Both CDM and SIDM with a large cross-section fail to reproduce
observed rotation curves of dwarf and LSB galaxies. We have seen that
the final density profiles are sensitive to the shape of the initial
fluctuations: more violent collapses end up with constant density
cores.  Alternatively, SIDM with a mean free path between kiloparsec
and megaparsec scales may solve this problem (Spergel \& Steinhardt
1999).  In this case, particles could transfer heat to the cold
central regions that occur in standard CDM collapses, creating an
initial expanding phase with lower central density. It is not obvious
that a cold core would be generated and maintained in a hierarchical
scenario since the dense mini-halos collapsing at high redshift may
form singular isothermal structures. The dense substructure halos
would rapidly sink to the centres of the parent halos by
hydro-dynamical drag, depositing high density low entropy material 
and conserving isothermal profiles.

Simulating intermediate mean free paths is relatively straightforward.
One technique would be to use the neighbour lists to choose random
particles to collide (Burkert 2000).  Simulations in progress will
demonstrate whether SIDM can reproduce the observed rotation curves of
dwarf galaxies. A solution to this problem will naturally resolve the
abundance of dark matter substructure in the Galactic halo since 
substructure with shallow potentials would be easily disrupted.

\acknowledgments

BM would like to thank Marc Davis for many discussions of the
astrophysical consequences of strongly interacting dark matter while a
NATO fellow in Berkeley and the Royal Society for support. VQ is a
Marie Curie research fellow (grant HPMF-CT-1999-00052). Computations
were carried out as part of the Virgo consortium.

\end{document}